\documentclass[a4paper,author]{erae}
\usepackage[T1]{fontenc}
\usepackage{times}
	\usepackage[utf8]{inputenc}
	\usepackage{csquotes}
	\MakeInnerQuote{"}
\usepackage[pdftex]{graphicx}
\usepackage{xcolor}
\usepackage{multirow}
	\usepackage{eurosym} 
\usepackage{tikz,amssymb,graphicx}
\usetikzlibrary{fit}

\usepackage{longtable}

	\usepackage{textcomp}
	\usepackage{amsmath,amsfonts}
	\usepackage{mhchem}
	\usepackage{multido}

	\usepackage{hyperref}
	\hypersetup{%
		 colorlinks = {true},
		 urlcolor = {black},
		 linkcolor = {black},
		 citecolor = {black},
		 pdfauthor = {Martial Phélippé-Guinvarc'h},
		 pdftitle = {Actuarial Implications of Yellow Virus in Sugar Beet After the EUs Ban on Neonicotinoids and Climate Change },
		 pdfkeywords = {Yellows Virus, actuarial model, Beet}
	}

\DeclareUnicodeCharacter{20AC}{\EUR{}}

\title{Actuarial Implications and Modeling of Yellow Virus on Sugar Beet After the EU's Ban on Neonicotinoids and Climate Change}

\author{Martial Ph\'elipp\'e-Guinvarc'h and Jean Cordier}

\date{Received: / Accepted: }

\begin{document}

\maketitle
%\tableofcontents

\begin{abstract}

Following the EU’s decision to ban neonicotinoids, % used to coat sugarbeet seeds,
 this article investigates the impacts of yellow virus on sugar beet yields under  the ban and under current and future climates. 
Using a model that factors in key variables such as sowing dates, phenological stages, first aphid flight and aphid abundance, simulations are performed using long-period climate datasets as inputs.
Coupled with  incidence and sugar yield loss assumptions, this model allows to reconstruct the impact of yellow virus on sugar beet yields using a so called `as if' approach. 
By simulating the effects of viruses over a longer period of time, as if neonicotinoids weren’t used in the past, this methodology allows an accurate assessment of risks associated with yellow viruses, as well as impact of future agroecological mesures.

The study eventually  provides an actuarial rating for an insurance policy that compensates the  losses triggered by those viruses.
\end{abstract}

\section*{Introduction}

During the 1990s, neonicotinoids emerged as a promising solution to manage the Yellow Virus risks on the sugar beet, eventually leading to their widespread adoption within a few years. 
However, in 2013, the European Union (EU)\footnote{\href{https://food.ec.europa.eu/plants/pesticides/approval-active-substances/renewal-approval/neonicotinoids_en}{https://food.ec.europa.eu}} imposed restrictions for the use of neonicotinoids in order to safeguard honeybees, which eventually led to a ban on  the use of neonicotinoids (NNI)  in 2018.
However, the regulation stipulated the possible use of NNI in very specific and emergency cases. 
Those emergency conditions were met in 2021 and 2022 and a temporary derogation was granted for the use of NNI as a seed-coating substance on sugarbeets in some Member States. 
However, on 19 January 2023 though, the European Court of Justice declared temporary derogations for the use of NNI as a seed-coating substance no longer valid. 
As a consequence, the use of neonicotinoids - as a seed-coating substance - is now prohibited in the EU. 

The objective of this paper is to assess the impact of this ban on sugar beet crop yield risks in the context of climate change. 
Our research is partially inspired by the recent paper by \cite{Dewar2021}, which examines the combined effects of climate change and the EU ban on neonicotinoids, and focuses on the incidence of yellow virus on sugar beets. 
We aim to study not only the incidence of yellowing viruses but also the increase in crop yield risks for the French sugar beet industry.
According to the literature, the timing of early aphid flights and phenological stages are strongly affected by climate. 
Additionally, field observations prove the severity of beet crop yield losses is influenced by the phenological stage of beets at the time of virus inoculation (performed by infected aphids). 
As a result, the impact of yellow virus infections is dependent on weather conditions and we utilize agronomic assumptions based on historical and projected climatic data to estimate the evolution of beets losses.

The pure risk premium of yellow virus correspond to the average economic losses due to yellow virus computed on a certain period of time. 
Since the pressure triggered by viruses  greatly varies from one year to another, it should be estimated over a long period of time.
 However, the analysis of historical data is `biased' by the use of neonicotinoids. 
It was therefore necessary to reconstruct the effects of yellow virus on sugar beet yields over an extended period using a statistical approach called `as if'. 
This method allows for the modeling of hypothetical effects of yellow virus on beet yields  `as if' they had occurred without any neonicotoids and over a longer period, thus enabling a better assessment of associated risks.

\section{State of the art}

The European Union is the world’s leading producer of beet sugar, accounting for about 40\% of the total beet sugar production worldwide.
Producing\footnote{\href{https://cefs.org/wp-content/uploads/2022/04/CEFS-Statistics-2020-2021.pdf}{CEFS Statistics --	European Association of Sugar Manufacturers}} 14.2 million tons of beet sugar in  2020/21, about  
25 000 workers are directly employed for the production of sugar in the EU while
105 000 growers -- representing a total  beet acreage of 1.41 million hectares -- supply beets to EU sugar factories. 

European sugar beet crops may be affected by several different yellowing viruses, in particular two Polerovirus, the  \emph{Beet Mild Yellowing Virus} (BMYV) and the \emph{Beet Chlorosis Virus} (BChV), and a Closterovirus, the \emph{Beet Yellows Virus} (BYV). 
 Two more rare viruses are also mentioned in the literature: the \emph{Beet Mosaic Virus} (BtMV) and the \emph{Beet Necrotic Yellow Vein Virus} (BNYVV).
%\citeauthor{Palak1979} in Czechoslovakia (\citeyear{Palak1979}).

Field infection by a yellow virus occurs in three stages.
Initially, a winged aphid (mainly the \emph{Myzus persicae}) acquires the virus from a virus reservoir, such as weeds, spinach, or unharvested beets. 
These winged aphids then introduce the virus to a beet field, contaminating some plants. 
Finally, a secondary infection occurs with winged or wingless aphids.% (mainly the \emph{Aphis fabae}). 
An aphid acquires a virus from an infected beet during the primary infection and spreads the infection by colonizing neighboring plants.
The severity of attacks depends on the amount of these aphids and the distance of virus reservoirs to the beet fields while beet yield losses are also affected by  virus prevalence. 
Early and widespread attacks are more likely to occur after mild autumn and winter, which allow for the development a larger initial aphid population in the following spring.

Symptoms of  yellow virus infections in sugar beets include yellowish discoloration of the leaves and necrosis, forming yellow circles in the field. 
Infected beet crops generally suffer losses in sugar production, with \emph{Polerovirus} potentially decreasing sugar beet yield by up to 30\% and BYV by up to 49\% according to \cite{Smith1990}. 
The main aphids responsible for beet yellowing are the green peach aphid (\emph{Myzus persicae}) and the black bean aphid (\emph{Aphis fabae}).

%https://exaly.com/trends/Google-Scholar
%https://app.dimensions.ai/discover/publication?search_mode=content&search_text=yellow%20virus%20beet&search_type=kws&search_field=text_search

The neonicotinoids, also known as `neonics' or  `NNIs,' are active substances used as plant protection products to control harmful insects. 
The name originates from their chemical similarity to nicotine and their insecticidal properties. 
The first neonicotinoid was approved in the EU in 2005.
Neonicotinoid seed treatment are systemic pesticides, meaning they are taken up by plants and distributed throughout various plant parts such as leaves, flowers, roots, stems, pollen, and nectar. Unlike contact pesticides that remain on the surface of treated plant parts, neonics penetrate the entire plant's tissues.
These substances are highly toxic to invertebrates, particularly insects, while showing lower toxicity towards mammals, birds, and other organisms. 
Neonicotinoids target the central nervous system of insects, ultimately causing paralysis and death.
Additionally, neonicotinoids are commonly used in veterinary applications for tick control and in the form of flea collars for pets.

The UK's literature has dealt with yellow virus for a very long time.
One of the earliest studies was conducted by  \cite{Hull1947}. 
This research was performed in the early forties in the UK at Hackthorn Lines and at Rothamsted Experimental Station in Harpenden. 
Other major papers include \cite{Smith1939},   \citeauthor{Russell1962} [\citeyear{Russell1963}, \citeyear{Russell1962}],  \cite{Watson1975},  \cite{HARRINGTON1989} , \cite{Qi2004} and  \cite{Dewar2021}, as well as a  literature review of the UK’s new virus detection covering a period of 35 years, from 1980 to 2014, done by \cite{Fox2017}.

Our literature review focuses on the initial aphid flights that mark the beginning of the infection,  the incidence or the dynamics of infection, and the crop losses caused by yellow viruses. 
Our assumptions were greatly influenced by those various points. 
% \cite{Johnson1929} calcul de l'estimation du taux de sucre par échantillonnage, insufisant sur 10 betteraves. 

\subsection{Aphid first flights}
%\cite{Moehring2020}
Even if initial relevant papers do exist on this specific aspect \citep{Qi2004,Watson1975}, the relationship between the seasonal timing of \emph{Mysus persicae} flights and winter temperatures in the UK has been well documented by \cite{Harrington2010}.
In France, \cite{Guillemaud2003} explain the spatial heterogeneity of aphid populations by looking at temperatures in winter.

More recently, \cite{Hemming2022} use observations from the same RIS suction-trap network than \cite{Harrington2010}. 
% à revoir: clarifier la phrase
Their model establishes the reliability between the day-of-year when 5\% of the total annual catch was observed on the one hand and the climatic data on the other hand. 
They confirm that aphids are coming earlier given the current climate change.
The recent work of  \cite{Luquet2023b} is relevant for our study because it was based on aerial aphid activity data collected by a French network of suction traps  that operated for more than 30 years. 

\subsection{Incidence}

The concept of incidence as well as models based on differential equations to describe infection dynamics have become more prevalent due to the COVID-19 pandemic.

Originally, the forecasting system developed by \citeauthor{Watson1975} (\citeyear{Watson1975}) allowed to predict, by the end of March, the extent of virus symptoms at the end of August. 
More recent papers such as \cite{HARRINGTON1989} and \cite{Qi2004} developed the same approach and helped the UK to implement a rational use of neonicotinoids based on early and later forecasts.

\cite{Werker1998} adapted the comprehensive incidence model developed by \cite{Brassett1988} to incorporate the specific characteristics of yellow virus infections. 
This model encompasses both primary and secondary infections.
The author suggests a model that relies on solving two interconnected differential equations: one for primary infection and the other for secondary infection. 
The model estimates based on UK regions requires the input of aphid abundance during the first flight.
We provide a detailed explanation of this significant model in a dedicated subsection.

Literature also highlighted another phenomenon: the more mature the beet is, the slower the aphid spread will be.
Recent research conducted by  \cite{Schop2022} supports and provides a comprehensive analysis of the influence of beet maturity on aphids. The resistance exhibited by mature sugar beet plants not only affects the mortality rate of Myzus persicae and the formation of a black deposit in their stomachs but also influences aphid fecundity and behavior.

%\cite{Yilmaz2015} virus proche dans les sols turques
\subsection{Yellow virus losses}

The challenge of accurately assessing losses due to yellow virus infections in the field is a significant concern for both practitioners and researchers.
Even today, French insurers continue to exclude health risks from Multi-Peril Crop Insurance policies. 
It is therefore crucial to distinguish between the amount of losses caused by disease and those caused by climate-related hazards when looking at losses. 
Similarly, in this paper, we examine the value of an insurance policy specifically targeting the yellow virus, which would require this distinction.

%The study's findings revealed that as yield per acre improved with better fertilization, the loss of sugar per acre caused by the virus also increased proportionally.
More recent field experiments  studied beet and sugar yield losses.
The Polerovirus (BChV, BMYV) and Closterovirus (BYV)  were already suspected to have different effects on yield by \citeauthor{Russell1962} (\citeyear{Russell1962},\citeyear{Russell1963}).
\cite{Smith1990} and \cite{Stevens2004} tested losses depending on the date of inoculation and the virus family.
They showed that early or mid-season inoculation could decrease sugar yields by up to 47\% for BYV,  29\% for BMYV, and  24\% for BChV. % (although the effect on sugar yield were more variable).
Experiments performed by  the \emph{`Institut Technique de la Betterave'}  (ITB)  revealed similar consistent results in France \citep[for detailed results]{ARTB2022}.
These papers also establish that losses decrease as beet maturity increases. 
In the case of co-infection,  Maupas and Brault (Jan 2022)\href{https://www.itbfr.org/tous-les-articles/article/news/impacts-de-la-multi-infection-de-la-betterave-sur-la-gravite-de-la-maladie/}{www.itbfr.org}  explain that there is no cumulative effect of the different viruses on the loss rate. The loss rates obtained are not significantly different from the loss rates of the most severe yellows.

The recent paper by \cite{Hossain2020}  assesses crop yield losses on a large scale. 
Leaf samples exhibiting virus-like symptoms were specifically gathered from sugar beet fields in 10 European countries over a period of 3 years. 
In total, over 6,000 samples were collected. 
Contrary to previous studies, \citeauthor{Hossain2020} showed that Polerovirus causes the most losses (23\% and 24\%) while a more moderate reduction in yields is observed for BYV (10\%). These results are surprising because they differ from all other and previous studies.

%\cite{CLOVER2001} test la croissance de la betterave en fonction de la evapotranspiration et de la radiation (en cas de virus ou d'absence de virus.)

In a scientific experiment such as the one conducted by \citeauthor{Stevens2004} (\citeyear{Stevens2004}), a plot of inoculated beets were compared to a plot of uninfected beets. 
However, outside of controlled experiments, it can be challenging to accurately determine the proportion of loss caused by the yellow  virus, especially when other damage occurs at the same time with similar symptoms (e.g. droughts) at the same time. 
This difficulty increases the risk of overcompensating or damaging the customer relationship due to inadequate or misunderstood indemnity assessments.
To address this issue, the implementation of an insurance policy should include an expertise protocol capable of distinguishing losses caused by the climate from those caused by the yellow virus. It would be valuable to consider the review by \citeauthor{Abade2021} (\citeyear{Abade2021}) on plant disease recognition using convolutional neural networks applied to images.

\section{Actuarial model}

In the present section, we establish a set of assumptions that link climate conditions to the rate of yellowing viruses losses.
These assumptions cover aspects such as incidence, phenological stages, probable sowing dates, crop areas, and beet loss rates in case of infection. 
We will apply these assumptions to multiple sets of climatic data for each agricultural region and each year, from which we will derive annual yellow virus loss rates.
The insurance policy is straightforward in its definition.  
The indemnity specifically provides targets losses caused by a yellow virus and covers the rate of losses (without deductible) multiplied by the expected price and expected crop yield.

\subsection{Aphid first flights assumptions}

Qi's studies (\citeyear{Qi2004,Dewar2021}) provide specific parameter sets derived from the UK. 
In both papers, the date of the first aphid flight is estimated using a linear regression of Growing Degree Days (GDD) between January 1st and February 14th.
The recent work of  \cite{Luquet2023b} was based upon aerial aphid activity data collected by a French network of suction traps  that operated for more than 30 years.
Because it uses the same method, the comparison was more interesting and  relevant.
%As precise by \citeauthor{Luquet2023b}, "this strategy of risk prediction and mitigation assumes a strong correlation between aphid flight features and disease expression in the field because it ignores the viruliferous status of aphids."

% latex table generated in R 4.2.1 by xtable 1.8-4 package
% Mon Jul 10 14:06:57 2023
Based on French observations, we tested the assumptions proposed by  \cite{Luquet2023b}. 
The first scenario, referred to as $M1-D1c$, utilizes the same dates as Qi's study. 
The second scenario, labeled $M2a-D1c$, incorporates an optimal range of dates. 
Additionally, the $Log_AbQIf$ model estimates aphid abundance based on the findings from Qi's research.
It is mainly observed that higher winter temperatures lead to earlier and more abundant aphid arrivals. 
What is even more surprising is that the temperature coefficients are similar despite the fact that the calculation periods for GDD are doubled between $M1-D1c$ and $M2a-D1c$ models.
\begin{center}
	\footnotesize
	\begin{tabular}{rlll}
			\hline
			& $M1-D1c$ & $M2a-D1c$ & $Log(Ab) QI_f$ \\ 
			\hline
			Intercept & 155.91 & 195.6 & -2.263 \\ 
			T crit & -0.1511 & -0.156 & 0.0423 \\ 
			critT start (dd/mm) & 01/01 & 01/04 & 03/12 \\ 
			critT end (dd/mm)& 02/14 & 03/28 & 03/29 \\ 
%			RSq\_(p) & 0.42 & 0.59 & 0.37 \\ 
			\hline
		\end{tabular}
\end{center}

\subsection{Incidences assumptions}

The model of \citeauthor{Werker1998} (\citeyear{Werker1998}) is our reference model (see also \cite{Gilligan1997}).
The observed proportion increase over time  of plants exhibiting symptoms of yellow virus infection (denoted as $Y$) can be attributed to two sources of inoculum, namely primary infection ($P$) and secondary infection ($S$). This increase is a result of the interplay between the proportion of disease-free plants ($1-Y$) and these two sources of infection.

Mathematically, this relationship is described by the following differential equations:
\begin{itemize}
	\item The rate of change of the proportion of plants with primary infection, denoted as $Y_{\mathrm{p}}$, with respect to time is given by 
	$dY_{\mathrm{p}}/dt = r_{\mathrm{p}}  P  (1-Y)$.
	\item The rate of change of the proportion of plants with secondary infection, denoted as $Y_{\mathrm{s}}$, with respect to time is given by 
	$dY_{\mathrm{s}}/dt = r_{\mathrm{s}}  Y  (1-Y)$.
\end{itemize}
Here, $Y_{\mathrm{p}}$ represents the proportion of plants with primary infection, $Y_{\mathrm{s}}$ represents the proportion of plants with secondary infection, $r_{\mathrm{p}}$ and $r_{\mathrm{s}}$ are rate constants for primary and secondary infection respectively.

Considering that the total proportion of diseased plants is given by $Y = Y_{\mathrm{p}} + Y_{\mathrm{s}}$, the combined rate of change of the total proportion of diseased plants with respect to time is given by 
$$dY/dt = (r_{\mathrm{p}}  P + r_{\mathrm{s}}  Y)  (1-Y)$$

In this investigation the infection cycle is initiated at time $t_0$,  then $Y_0=0$ and the solution are:

\begin{equation}\label{Y}
	Y=\frac{1-e^{-\left(r_{\mathrm{p}} P+r_{\mathrm{s}}\right)\left(t-t_0\right)}}
	{1+\frac{r_{\mathrm{s}}}{r_{\mathrm{p}} P} e^{-\left(r_{\mathrm{p}} P+r_{\mathrm{s}}\right)\left(t-t_0\right)}} 
\end{equation}
Time $t_0$ coincides with the start of the spring migration of the aphid species Myzus persicae, and at this point, the proportion of infected plants, denoted as $Y_0$, is assumed to be zero.

The primary inoculum, represented by the variable $P$, is defined as the number of migrant viruliferous aphids that land on the crop and feed on it. The determination of $P$ is based on the total number of aphids migrating in the spring, which is assessed using suction traps from the Rothamsted Insect Survey. The relationship between $P$ and $N$, the total number of spring migrating aphids, is established by the equation:
\begin{equation}\label{N}
	P=1 - e^{-pN}
\end{equation}
This equation corresponds to the multiple infection transformation and involves the constant of proportionality denoted as $p$. The use of this transformation is necessary to stabilize the values of $P$ since the number of migrating aphids ($N$) exhibits significant variations from one year to another.

The difficulty lies in determining the relevant parameters for this model based on UK's suction traps network.
 The literature from the UK offers certain sets of parameters, but we observe variations among different regions within the UK (see \cite{Dewar2021}, \cite{Qi2004} and \cite{Werker1998}).  
Then, we retain these parameters as assumptions in our model: the $QiN$  being the most suitable, according to  ITB beet experts.

\begin{table}[htp]
	\begin{center}
		\footnotesize
		\begin{tabular}{rrrrr}
			\hline
			& Werker98 & $QiE$ & $QiN$ & $QiW$  \\ 
			\hline
			$p$ & 0.0562 & 0.01095 & 0.01276 & 0.007514  \\ 
			$r_{\mathrm{p}}$ & 0.0001 & 0.00205 & 0.00250 & 0.00112  \\ 
			$r_{\mathrm{s}}$ & 0.0409 & 0.06920 & 0.08440 & 0.04  \\ 
%			$\gamma$ & 0.00 & 0.01 & 0.01 & 0.01 \\ 
			\hline
		\end{tabular}
	\end{center}
	\caption{Incidence's set of parameters found in literature}
\end{table}

\subsection{Phenological stages and losses assumptions}

\begin{description}
	\item[Seedling]

We have not found any literature on a model for the evolution of planting in the agricultural crop year, neither for sugar beets nor for other crops. 
Therefore, our assumptions reflect the expertise expressed by ITB.

The weather conditions must be favorable for germination and for the circulation of a tractor equipped with a seeder. For example, a field that is too wet becomes muddy, therefore preventing tractors to go in field. 
Seeding is only feasible if the rainfall before and after seeding is not excessive and if the cold weather is not extreme during that period. Sometimes, planting can be done over consecutive days, while other times it may only be feasible within a day or two, resulting in fragmented planting periods.

A favorable day is considered to allow for a constant percentage of the total sowing. However, factors such as wind and evapotranspiration, which could potentially decrease the impact of rainfall and increase the number of suitable days, are not taken into account.

Seedling is carried out as soon as possible after March 10th and before May 10th. Planting is feasible if the minimum temperature is above -3°C from the day before (j-1) to three days after (j+3), if the accumulated rainfall over the past five days does not exceed 4 millimeters, and if the anticipated rainfall for the day of planting and the following two days does not exceed a cumulative 5 millimeters.

If we assume that seven feasible dates are needed to reach 100\% of planting (around 15\% per day), our results are generally consistent with experts from ITB.
The discrepancy between simulated and observed results is particularly noticeable for late plantings. 
However the increasing pressure of yellow virus may lead to a reduction of this late sugar beet plantings. As a result, our simulated results are likely to provide more reliable information than historical data, especially regarding the tail end of the distribution.
Our approach remains original as it takes into consideration the wide range of possible planting dates, which strongly affects the vulnerability of the plant.

	\item[Phenological stages]
Agronomists describe the phenological stages of sugar beets based on the number of leaves. These stages are highly dependent on temperature accumulation and were modeled using a set of parameters provided by ITB: 
$$
\ell=  -3.0834 + 0.019734 \times GDD
$$
where $\ell$ estimated the number of leaf and $GDD$ represents the cumulative growing degree day with threshold 0, right after seedling is performed.
To simplify matters, our assumptions regarding the stages are directly expressed in Growing Degree Days (GDD, see following table).
	
\item[Prevalence:] To date, the literature does not provide well-documented estimations of future prevalence after the ban of neonicotinoids and under the influence of climate change.
ITB  recorded a prevalence of 7\% for BYV in France in 2019, followed by 70\% in 2020, 35\% in 2021, and 68\% in 2022. These data present significant variations over time. 
Similarly, \cite{Hossain2020} confirm the presence of strong heterogeneity across Europe during the period from 2017 to 2019.
Although co-infection is common and this article deals with co-inoculation, it does not provide statistics on co-infection or multi-infection.

Due to this substantial variability in virus prevalence, it becomes challenging to choose relevant assumptions. 
Therefore, we  conducted two separate evaluations, one for Polerovirus (BMYV/BChV) and another one for the strong yellow virus (BYV) that triggers the most significant beet yield losses. 

	\item[Losses] The literature and tests conducted by ITB consider loss rates based on the combined dates of inoculation and  plant development stage. 
	This led us to propose yield loss rates established in the following table.\footnote{
		\citeauthor{Russell1962} [\citeyear{Russell1963}, \citeyear{Russell1962}] give effect of  inoculation at the stage of 4-6 leaves with BYV of 40\%-47\% (1985 Trial A-C) and of 22\% (1987)  loss and with BMYV of 27\%-29\% (1985 Trial A-C) and of 27\% (1987).
		\cite{Smith1990}  give effect of  inoculation at the end of May with BchYV of 8, 20, 24\% loss (1997, 1999 \& 2000)  and with BMYV of 22, 27 \& 27\%. 
		ITB's experimentation  obtains a loss of 52\% for BYV and 27\% for BMYV/BchyV (14\% for BtMV - Polyvirus)\citep[for details]{ARTB2022}.
	} 
To ensure consistency, we utilize a spline to interpolate intermediate dates between the reference dates of inoculation.
 In the case co-infection,  we only consider the most severe infection according to results of Maupas and Brault (Jan 2022).
	
	\begin{center}\footnotesize
		\begin{tabular}{lrrr}\hline
			\textbf{Stage} &$GDD$ &\multicolumn{2}{c}{\textbf{ Loss rate}} \\
			 & &\textbf{BMYV/BChV}& \textbf{BYV} \\ \hline
			Emergence &180 & & \\ 
			4-6 leaves &400 & 30\%& 50\%\\
			12 leaves &765 & 19\%& 29\%\\
			18 leaves &1070 & 11\%& 31\%\\
			Maturity &1200 &3\%& 23\%\\
			\hline
		\end{tabular}
	\end{center}

It is important to note that the article by \cite{Hossain2020} contradicts other elements in the literature. The reported losses are significant, with a loss of 29.8\% for BMYV or BchV, 11\% for BYV, and 45\% for coinfection BchV/BYV. 
Despite this contradiction, it is interesting to observe that different levels of loss rates are similar, even though they do not pertain to the same virus.

\end{description}

\subsection{Climate open data}

DRIAS DATA is a comprehensive dataset that includes observed climate data (1950--2005) and climate model projections (2006-2100) developed by a collaborative effort involving Météo-France, CERFACS\footnote{CERFACS stands for "Centre Européen de Recherche et de Formation Avancée en Calcul Scientifique," which translates to "European Centre for Research and Advanced Training in Scientific Computing" in English. It is a research institution located in France that specializes in advanced scientific computing and simulation, particularly in the fields of climate, weather, and environmental sciences.}, IPSL\footnote{IPSL: Institut Pierre-Simon Laplace is a French research institute focused on climate science.}, and others, noted 'EXPLORE2-Climat 2022'. It provides reliable and up-to-date climate information with various variables like temperature, precipitation, wind speed, and more. The dataset offers projections based on different emission scenarios known as representative concentration pathways (RCPs) like RCP2.6, RCP4.5, RCP6.0, and RCP8.5. DRIAS DATA uses a spatial grid system to organize climate data at different geographic locations.
In this paper, we choose to compute the model for each climate model and RCP combination and then average the results for each RCP.

\cite{Kapsambelis2022a} %\cite{Kapsambelis2019,Kapsambelis2022} 
utilizes a single climate model, namely the ARPEGE-Climat model. This particular model stands out as it encompasses a broad range of simulations under consistent climate conditions while incorporating climate change. 
The reliance on a single model raises concerns about the reliability of the obtained results. 
In order to address this limitation, the author choose to compare their findings with multiple models from the "EURO-CORDEX" project (2014). 
Their analysis reveals that taking the mean of multiple models introduces biases and underestimates extreme values.
Although well-suited for our study, these ARPEGE data are not accessible to us. 
We can hope that between the 11 models provided by DRIAS in 2014 and the current 20 models, the corrections made will mitigate the problem highlighted by \cite{Kapsambelis2022a}.

%{\color{blue}
%JRC MARS climate data, part of the Agri4cast project, is provided by the JRC of the European Commission. It focuses on climate variables important for agricultural applications, such as temperature, precipitation, solar radiation, wind speed, and humidity. The data is collected and processed into comprehensive climate datasets covering multiple European regions (1972--2021 at writing date of this paper).
%While DRIAS primarily focuses on France and offers detailed regional-level information (Grid 8km x 8km), Agri4cast climate data covers recent years with real data (Grid 25km x 25km).}

\subsection{Surface, crop expected yields and prices}

\paragraph{The graphic land registry database:}
The graphic land registry database (RPG - 2021)  has been provided by the service and payment agency (ASP) since 2007 (\href{https://geoservices.ign.fr/documentation/diffusion/telechargement-donnees-libres.html\#rpg}{geoservices.ign.fr}).
It is used to calculate public European Common Agricultural Policy (CAP) aids, with greater reliability than a survey. 
It offers an incomparable level of geographical detail on sowing.

\paragraph{The Farm Accountancy Data Network:} FADN is a database that contains accounting information from individual farms.  
The FADN database focuses on EU agricultural holdings that can be considered `commercial' based on their size, as explained by the European Commission.  
The FADN data offers several advantages, including a long historical dataset (1988-2022), a weighted sample built using sampling rules, and controlled results verified by a chartered accountant.
In France, it is referred to as RICA (Réseau d'Information Comptable Agricole) and has been collecting data since 1989. 
 The chosen sample consists of 7,600 French farms, representing approximately 300,000 farms in France. 
 The database includes the income statement and balance sheet for each farm, as well as additional information such as farm structures, crop-growing practices, and production results which are collected through an annual survey. 
 While the data does not directly provide crop yields, they do include information on crop production and acreage, which therefore allows for yield calculations.
 The expected crop yields for each agricultural region were calculated based on the five-year Olympic mean.
%
%
%We have acquired data from two private institutions. Firstly, ITB has provided a sample of 10 farms covering the period from 1989 to 2021. This dataset has been enriched with observations related to yellow virus and includes detailed harvest results such as Potassium meq, Sodium en meq, Azote en meq, Glucose en meq, Poids racine (T/ha), and Rendement à 16° (T/ha).  This data is obtained through field sampling, ensuring a direct link between the field and the reported results. However, it's important to note that there may be some measurement errors associated with this method.
%
%Secondly, Tereos has provided comprehensive data from their cooperative members, covering the period from 2001 to 2021. The farm information, including names and locations, has been anonymized in the dataset. This dataset encompasses over 11,000 beet growers affiliated with the cooperative. This large dataset that is statistically significant. This comprehensive dataset covers a wide range of beet growers and offers valuable insights into overall trends and patterns. Unlike ITB, Tereos establishes crop yield based on delivery measurements, which provides a reliable measure of the actual crop production. However, this method does not guarantee a direct link between the field conditions and the reported measurements.

\paragraph{Price:} The price of sugar beets is assumed to be constant in our model. We set it at the current price of 45€ per ton at a standardized 16\% sugar content.

\section{Results}

The `as if' model was run using 92 different configurations. 
The tests referred to climate simulations from six institutions, each providing two versions of the model. 
These multiple scenarios were developed to stabilize and control results. For our tests, we selected two scenarios from the IPCC, namely Scenario 4.5 (20 configurations) and Scenario 8.5. We also considered two modalities for the first aphid flight and two modalities for the incidence model. 
By incorporating multiple possible assumptions, we were able to explore the sensitivity of our model to the chosen hypotheses.
Unsurprisingly, each model shows an increase in losses as the climate evolves, highlighting two challenges that farmers will face: the absence of neonicotinoid insecticides  and climate change. 

The total area considered in our study spans 419,499 hectares across the 429 agricultural regions. Based on current prices of 45€/t, the expected production -- which is estimated to be 33.6 million tons -- represents a total value of 1.512 billion euros.
When analyzing various scenarios from 1950 to 2100, the average loss for France is estimated to be 7.3\%, equivalent to 110 million euros. However, it is important to note that averaging these simulations may not provide meaningful insights. For instance, 23 configurations (25\%) utilize the incidence model parameterized in the eastern region of the United Kingdom ($QiE$). 
Nevertheless, expert tests conducted by ITB have determined that the most suitable incidence model is the one parameterized in the northern region of the United Kingdom ($QiN$), as presented in Qi's article. 
Despite these variations, the simulations performed allowed us to evaluate the sensitivity of this assumption within the model.

%Table \ref{Statdes} offer detailed descriptive statistics of our 92 simulations results.

%\include{Statdes} % Statistiques descriptives

To facilitate the understanding of results through the different configuration runs, we computed a linear regression on data results.
In these, the reference is the result of Institute CNRM-CERFACS-CNRM-CM5\footnote{Acknowledged as one of the leading international lab for meteorological research, CNRM is not only the main Météo France lab in R\&D but also synchronizes all Météo France R\&D in an Earth system model designed to run climate simulations. It consists of several existing models designed independently and coupled through the OASIS software developed at CERFACS.}, the projection GIEC 8.5 and the models of first flight $M1-D1c$ and of incidence $QiN$ recommended by ITB experts.

\begin{table}[htp]
\begin{center}\footnotesize
\begin{tabular}{l c c c c c}
\hline
 & Frist Model  &  \multicolumn{4}{c}{Without Institution variable}   \\
 & all years & all years & current years	&  future years &  far years \\ 
& 1950--2100   & 1950--2100 & <2025 & $\geq$2025 & $\geq$2050 \\
\hline
(Intercept)                           & $0.0426^{***}$  & $0.0420^{***}$  & $0.0238^{***}$  & $0.0588^{***}$  & $0.0686^{***}$  \\
                                      & $(0.0038)$      & $(0.0024)$      & $(0.0015)$      & $(0.0036)$      & $(0.0042)$      \\
Institution ICHEC-EC-EARTH    & $0.0009$        &                 &                 &                 &                 \\
                                      & $(0.0041)$      &                 &                 &                 &                 \\
Institution IPSL-IPSL-CM5A-MR & $0.0380^{***}$  &   \multicolumn{4}{c}{ \emph{rows of model IPSL exclude} }               \\
                                      & $(0.0041)$      &                 &                 &                 &                 \\
Institution MOHC-HadGEM2-ES   & $-0.0019$       &                 &                 &                 &                 \\
                                      & $(0.0045)$      &                 &                 &                 &                 \\
Institution MPI-M-MPI-ESM-LR  & $-0.0108^{*}$   &                 &                 &                 &                 \\
                                      & $(0.0041)$      &                 &                 &                 &                 \\
Institution NCC-NorESM1-M     & $-0.0049$       &                 &                 &                 &                 \\
                                      & $(0.0041)$      &                 &                 &                 &                 \\
Scenario rcp45                       & $-0.0146^{***}$ & $-0.0134^{***}$ & $-0.0012$       & $-0.0242^{***}$ & $-0.0318^{***}$ \\
                                      & $(0.0036)$      & $(0.0032)$      & $(0.0020)$      & $(0.0048)$      & $(0.0057)$      \\
Virus BYV                              & $0.0563^{***}$  & $0.0518^{***}$  & $0.0313^{***}$  & $0.0707^{***}$  & $0.0804^{***}$  \\
                                      & $(0.0024)$      & $(0.0022)$      & $(0.0013)$      & $(0.0033)$      & $(0.0039)$      \\
Model Pucerons M2a-D1c                 & $0.0328^{***}$  & $0.0285^{***}$  & $0.0042^{*}$    & $0.0509^{***}$  & $0.0634^{***}$  \\
                                      & $(0.0034)$      & $(0.0030)$      & $(0.0019)$      & $(0.0046)$      & $(0.0053)$      \\
Model Diffusion QiE                         & $-0.0261^{***}$ & $-0.0249^{***}$ & $-0.0166^{***}$ & $-0.0325^{***}$ & $-0.0362^{***}$ \\
                                      & $(0.0034)$      & $(0.0030)$      & $(0.0019)$      & $(0.0046)$      & $(0.0053)$      \\
\hline
R$^2$                                 & $0.9275$        & $0.9266$        & $0.9063$        & $0.9233$        & $0.9241$        \\
Adj. R$^2$                            & $0.9195$        & $0.9224$        & $0.9010$        & $0.9189$        & $0.9198$        \\
Num. obs.                             & $92$            & $76$            & $76$            & $76$            & $76$            \\
\hline
\multicolumn{6}{l}{\scriptsize{$^{***}p<0.001$; $^{**}p<0.01$; $^{*}p<0.05$}}
\end{tabular}
\caption{Statistical results of simulations}
\label{table:coefficients}
\end{center}
\end{table}

We observe in table \ref{table:coefficients} that the Climate Institution is less significant than other variables. With the exception of the climate scenario proposed by the French institution IPSL, the choice of model does not strongly affect the results. At times, we do not have an explanation for the specificity of this scenario.

For the reference assumptions and for all climate institution except IPSL we obtain the 20 simulated configurations, ten per virus type.
For all years (1950--2100), the average loss is estimated at 4.04\% (BMYV/BchV) and 9.55\% (BYV), equivalent to 61 and 143 millions of euros.
Unfortunately, the average losses are not meaningful because the loss series are not stationary. The change in climate is in our model the main reasons for the non-stationarity of the `as if' loss series.

If we consider only the current period (before 2025), the average loss is estimated at 2.21\% (BMYV/BchV) and 5.69\% (BYV), equivalent to 33 and 86 millions of euros. 
For the futures years  (2025--2100), considered as the future climate, the average loss is up to 5.72\% (BMYV/BchV) and 13.11\% (BYV), equivalent to 86 and 197 millions of euros. 
For the fast years  (2050--2100), considered as the far climate, the average loss is up to 6.70\% (BMYV/BchV) and 15.07\% (BYV),  equivalent 101 and 227 millions of euros.
Then, climate change is significantly increasing future losses and yellow viruses could cost european farmers 227 millions of euros per year in the future.

\begin{center}\footnotesize
	\begin{tabular}{l c c c c c}
		\hline
		&  & all years & current years	&  future years &  far years \\ 
		& \textbf{Unit} & 1950--2100 & <2025 & $\geq$2025 & $\geq$2050 \\
		\hline
		\textbf{BMYV/BchV} &\textbf{\%} & 4.04\%  & 2.21\%  & 5.72\%  &  6.70\% \\
		&\textbf{M€}& 61      & 33    & 86      & 101    \\\hline
		\textbf{BYV} &\textbf{\%}& 9.55\% & 5.69\%   & 13.11\%  & 15.07\%    \\
		&\textbf{M€}& 143      & 86      & 197     & 227   \\\hline
	\end{tabular}
\end{center}

In our approach, the average values represent also the pure premium of an insurance policy that covers losses resulting from the yellow virus. This insurance policy assumes no deductible, and the loss rate due to the yellow virus is applied to the expected production value.

The other coefficients of regression  in table \ref{table:coefficients}  align with the expected outcomes. Scenario 4.5, which represents a less severe climate change hypothesis, yields a negative parameter, while the BYV virus, associated with severe yellowing, yields a positive coefficient. Both aphid flight models are parameterized using French data. We are facing a typical model risk, and this result suggests that the more cautious model, the M2a-D1c, should be adopted. Similarly, among the two incidence models, the one recommended by the technical institute corresponds to the more prudent option.

\begin{figure}[htp]
	\centering
	\includegraphics[width=0.8\linewidth]{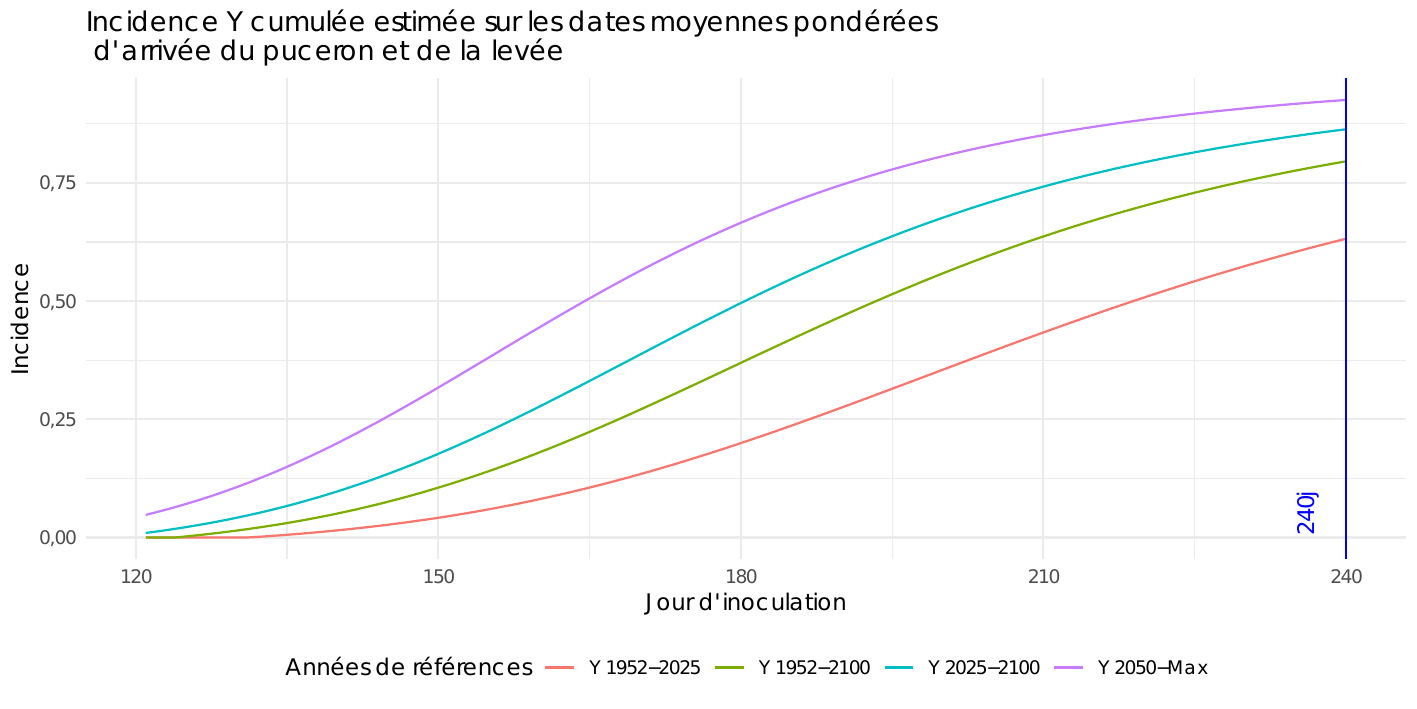}
	\caption{Incidence that depend on reference periods}
	\label{fig:incidence}
\end{figure}

In  Figure \ref{fig:incidence}, the difference between the four curves is striking.
It is obtained using the reference assumptions mentioned above, based solely on the CNRM climate model (Météo France). 
The further into the future, the stronger and earlier the rate of contamination becomes. 
This is true regardless of the type of virus, because the evolution of colonization by aphids does not depend on the virus it carries.
Changes in climate tend to bring forward the arrival of aphids and accelerate plant development, but not only that. It tends to increase the rate of contamination of sugar beet. 

\begin{figure}[htp]
	\centering
	\includegraphics[width=0.8\linewidth]{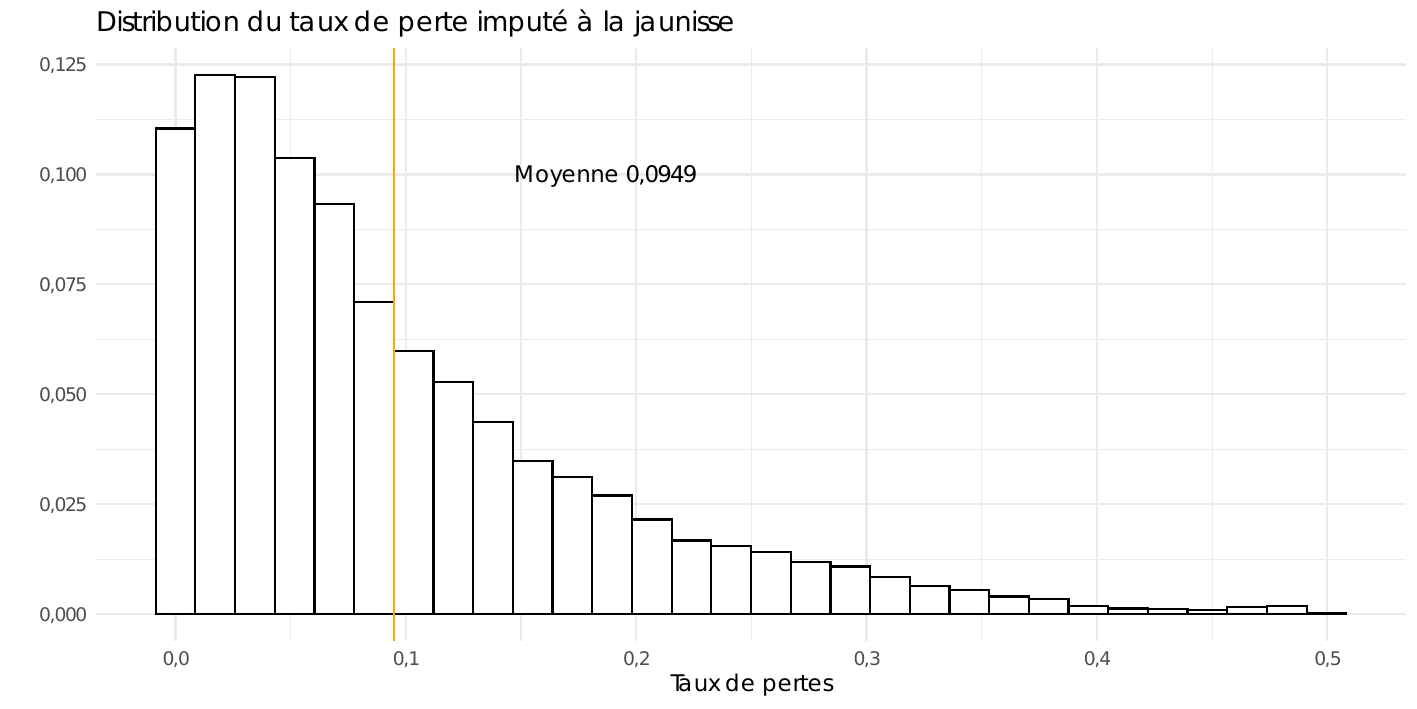}
	\caption{Histogram of  loss rates by agricultural regions on the reference configuration}
	\label{fig:tphisttx}
\end{figure}

\begin{figure}[htp]
	\centering
	\includegraphics[width=0.8\linewidth]{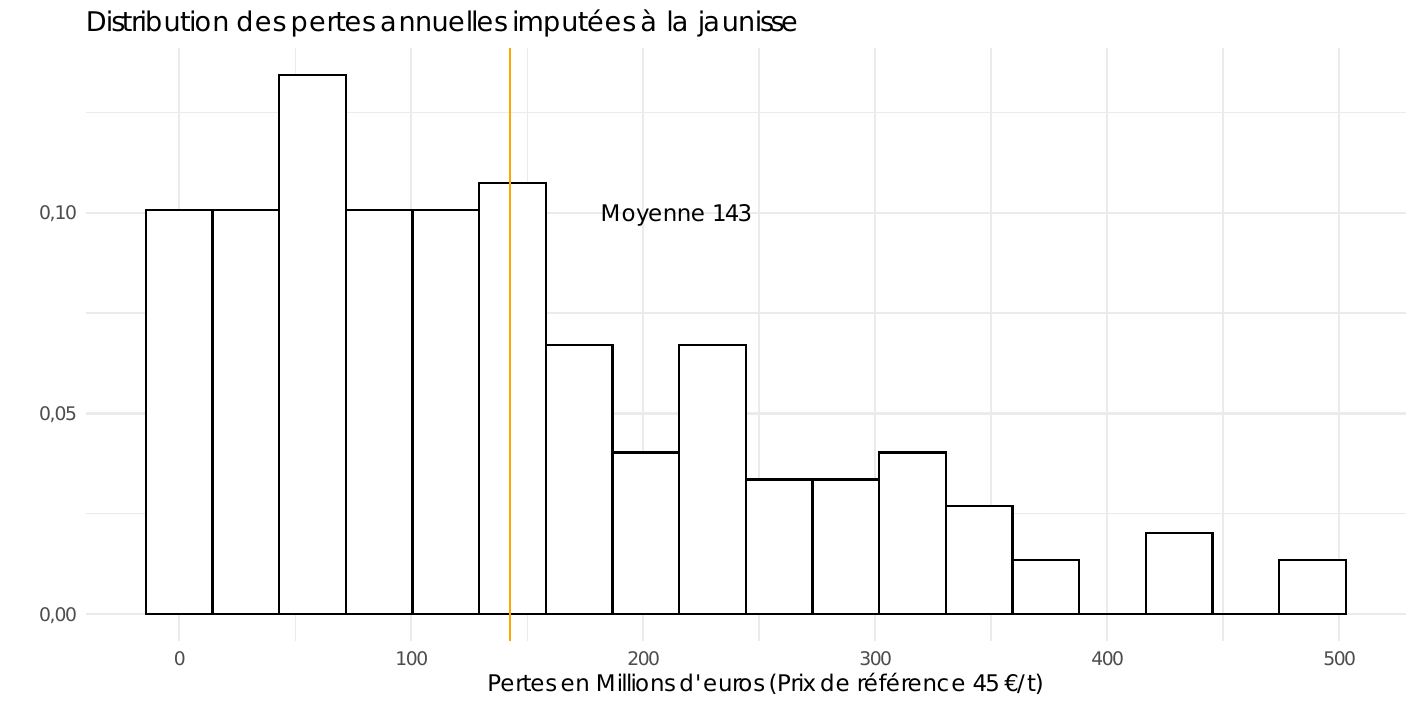}
	\caption{Histogram of  loss sums by year on the reference configuration}
	\label{fig:tphistsum}
\end{figure}

In Figure \ref{fig:tphisttx} and  \ref{fig:tphistsum}, only BYV is considered.
The Figure \ref{fig:tphisttx} illustrates the average loss rates for agricultural regions (weighted by RPG surface area) and the figure  \ref{fig:tphistsum} displays annual losses for the French farm, expressed in millions of euros.
These two graphs clearly demonstrate the variability in the risk associated with yellows even if a part of this variability is attributable to the upward trend in losses, as it encompasses all years.

The variability depicted in Figure \ref{fig:tphisttx} has the greatest impact on farmers. 
The higher the variability, the greater the risk of experiencing major losses. 
However, this variability does not necessarily have an impact on insurers if they have a sufficient number of underwritten  policies.

Figure  \ref{fig:tphistsum} is the most important for insurers. The pronounced dispersion shown in this graph indicates that the risk has a systemic component. Even with a large number of policies insured, intra-annual risk pooling remains partial, and insurers retain a substantial level of risk for themselves.
This result was expected.
Given the importance of climatic data in our model, we did indeed expect a strong spatial correlation and, consequently, a strong systemic component on the risk of yellow virus.

\section{Discussion}

With an estimated average future yield loss of around 10\%, the ban on neonicotinoids has a significant economic impact for farmers and the sugar industry. 
The reasons for a ban on neonics primarily stems from environmental concerns and they have sparked significant research interest. 
\footnote{For instance, a recent study by \cite{Odemer2023} examined the transfer of neonicotinoids to honey bees.}

Our work stands out because it considers economic losses due to a wide range of climatic, agronomic, and health related assumptions and values.
It is worth noting that many lab findings are yet to be validated under field conditions, and results from small-scale plot tests which may not necessarily reflect the outcomes on a larger scale. 
Disease development in plants is influenced by many factors, including the presence of a susceptible host, a virulent pathogen, and suitable environmental conditions. 
For example, \cite{Velasquez2018} have shown that, on climate variables alone, temperature, \ce{CO2} concentration, water availability (water stress) can affect plant resistance. 
Furthermore, models are commonly parameterized based on historical data, which introduces limitations in predictability. 
We have therefore taken great care in selecting these assumptions, consulting both relevant literature and agricultural experts.
Our results provide valuable insights into the complex factors at stake within the sugar beet ecosystem, although they may not capture every nuance of such complexity. 
While our projection model does not offer future predictions, it provides valuable guidance for farmers and insurers in adopting new practices.

In addition, it is important to highlight that the ban on neonicotinoids combined with climate change will not be the only developments. 
Several other factors will certainly play a role in shaping the future of sugar beet cultivation:
\begin{description}

\item[New viruses:]  The emergence of new viruses, such as the distinctive yellow leaf spot caused by the Stemphylium sp., has been observed in sugar beet cultivation, as explained by \cite{Hanse2015} in the Netherlands.
But this prospect remains as unpredictable as a COVID-19 pandemic, and unfortunately cannot be reasonably taken into account.

\item[New cultural practices in beet:] Farmers will adopt new cultural practices such as companion/ecosystemic planting practices, biocontrol solutions, and genetic improvements to adapt to a fast-moving environment.
Ongoing research is actively exploring these solutions that may enrich the findings of our current study.
 For example, \cite{Albittar2019} discusses the possibility of using the \emph{parasitoid L. fabarum} for targeted biological control against aphids, as it does not seem to be affected by the Yellow virus.
According to a recent study by \cite{Francis2022}, the implementation of innovative eco-compatible approaches, such as  the use of semiochemicals, Entomopathogenic fungi, and Plant Growth Promoting Rhizobacteria  in combination with the use of `resistant' beet varieties, show  promising results when comes to achieving sustainable pest management strategies.

The recent paper of \cite{Laurent2023} examines small-plot trial data and founds that products containing spirotetramat and flonicamid as active ingredients can significantly reduce aphid numbers.
When applied at the recommended rate without an oil-based adjuvant, spirotetramat and flonicamid achieved effectiveness rates of 70.6\% and 58.9\% respectively, 14 days after application. 
The effectiveness increased to 85.6\% and 79.9\% respectively when combined with an oil-based adjuvant. The study also highlighted the promising potential of a biopesticide based on Lecanicillium muscarium strain Ve6, which achieved 40.7\% effectiveness. 

In France, the National Plan for Research and Innovation (PNRI), initiated in January 2021, aims to explore alternative solutions to neonicotinoids. Under this plan, various products are being tested, including Lecanicillium, maltodextrin, and paraffin oil.
For instance, maltodextrin is derived from plant such as corn and undergoes a rapid drying process that forms an air-impermeable meshwork by cross-linking glucose polymers. This unique property enables maltodextrin to act as an insecticide by obstructing the spiracles of target insects, leading to suffocation and inhibiting their ability to fly.
In addition, the PNRI also investigates the efficacy of Volatile Organic Compounds (VOCs) in granulex form, the release of beneficial insects such as chrysopid eggs or larvaes, and the use of ecosystemic plants during the seedling stage. 

\item[Global changes:] 
The ban on neonicotinoids is part of a broader movement to phase out the use of synthetic pesticide  molecules in various agricultural activities. 
Legislation like the French law N°2014-110 named `loi Labbé', enforced in 2017, prohibits their use  in public spaces and non-agricultural activities. 
Recently, in February 2023, ANSES initiated a procedure to withdraw authorization for the use of plant protection products containing S-metolachlor to preserve groundwater quality. 
Additionally, a bill for zero net artificialization is currently under review. 

In addition to these regulatory changes, there are other challenges posed by climate change, agro-ecological transition, emerging technologies, and the demography shifts in the farming community.
This dynamic of change presents opportunities for further research on the evolution of farming practices, particularly with a view to the planned reduction of pesticides use and the effects of climate change, and this is likely to offer valuable insights to our study.

%\item[Potential ecosystem rupture:]  The impact of these changes extends beyond the ban on neonicotinoids and affects the broader ecosystem. While our study focuses on sugar beet, it is essential to recognize that changes in agricultural practices, new regulations, and the emergence of new pathogens have implications for all crops and the entire ecosystem.
%
%It is worth noting that outside the tropics, there is a global trend of increased prevalence of overwintering pathogen inocula, potentially leading to more severe and frequent epidemics \cite{Velasquez2018, Hallmann2022}. In agroecosystems, plant viruses can impact the interactions between crops, crop pests, and natural enemies. 

\end{description}

\section*{Conclusions}
%à corriger 
%parler dans les résultat de la variabilité correlation

In the absence of neonicotinoids, this article focuses on the specific impact of yellow virus on sugar beet yields. 
It employs a simulation approach that takes into account various factors, such as sowing dates, phenological stages, first aphid flight, and aphid abundance, using assumptions derived from relevant literature. 
These simulations utilize climate datasets as inputs.
By incorporating these assumptions, along with the incidence and loss assumptions, the study reconstructs the effects of yellow virus on sugar beet yields using the `as if' approach. This methodology allows for a more comprehensive assessment of the associated risks by modeling the virus's impact over an extended period, as if neonicotinoids were not used.

Using agronomic and climate assumptions, the study also provides an actuarial rating for yellow virus losses which, on average, amount to  approximately 10.2\%, or 154 million euros.
This study also highlights the unfavorable trend resulting from climate change as well as the significant annual variability of beet yellow losses. 
It also confirms the strong systemic component of this risk. 
As a result, the development of a yellow virus insurance will require reinsurance coverage, possibly through public reinsurance.

Practically and instead of creating a stand-alone yellow virus insurance, it was found to be more cost-effective and relevant to include a new yellow virus guarantee within an existing multi-peril crop insurance. 
Future work will focus on combining risk assessment and investigating the relationships between crop yields affected by climate and crop yields affected by yellow virus.

We are also keen to incorporate the impacts of emerging practices, such as ecosystemic  planting, biocontrol solutions, and genetic progress, into our simulations. 
Some emerging practices  are currently undergoing validation and development within the PNRI's program. 
This integration will enable us to evaluate the effectiveness of these innovative approaches and explore their synergistic effects with climate factors on crop yield outcomes. 
We expect that incorporating these new practices will improve the accuracy and relevance of our simulations, providing valuable insights for  agronomic and insurance decision-making.

\subsection{Acknowledgment}
Thanks 	to Luc Boucher, director of Risques Agricoles (DiagoRisk),  %Jean Cordier, Emerita Professor,
 to Azilis Lesteven  and  Alexis Patry, Association de Recherche Technique Betteravière (ARTB), 
 to Fabienne Maupas, ITB, and
 to Martin Luquet, INRAe, for their valuable comments and suggestions. 

\subsection{Funding}

The  National Research and Innovation Plan (PNRI) is financing 23 projects and mobilizes many different players. 
. This work was performed within the fourth axis of PNRI intituled ‘Transition towards a sustainable economic model’ as part of the \href{https://www.itbfr.org/pnri/projets/grecos/}{Grecos} project led by the Association de Recherche Technique Betteravière (ARTB) : a project which aims to prefigure what risk management system would be best suited given the alternative technical solutions developed by other project of the PNRI.

\bibliographystyle{erae}

\end{document}